\documentclass[twocolumn,final,balancelastpage,superscriptaddress,nofootinbib]{revtex4}

\usepackage{graphicx}

\usepackage{tabularx}
\usepackage{amssymb}
\usepackage{amsmath}
\usepackage{amsbsy}
\usepackage{comment}
\usepackage{mathrsfs}
\usepackage{dsfont}
\usepackage{nicefrac}
\usepackage{wrapfig}
\usepackage{amsfonts}
\usepackage{url} 
\usepackage[nolist]{acronym}
\usepackage{nicefrac}
\usepackage{subfigure}
\usepackage{epsf,color,colordvi,pifont}
\usepackage{showkeys}
\DeclareFontFamily{OT1}{pzc}{}
\DeclareFontShape{OT1}{pzc}{m}{it}{<-> s * [1.10] pzcmi7t}{}
\DeclareMathAlphabet{\mathpzc}{OT1}{pzc}{m}{it}

\usepackage{slashed}

\begin{document}
	\title{Impact of laser focussing and radiation reaction on particle spectra from nonlinear Breit-Wheeler pair production in the nonperturbative regime}
	\author{A. Eckey}
	\affiliation{Institut f\"ur Theoretische Physik I, Heinrich Heine Universit\"at D\"usseldorf, Universit\"atsstr. 1, 40225 D\"usseldorf, Germany}
	\author{A. Golub}
	\affiliation{Institut f\"ur Theoretische Physik I, Heinrich Heine Universit\"at D\"usseldorf, Universit\"atsstr. 1, 40225 D\"usseldorf, Germany}
	\author{F. C. Salgado}
	\affiliation{Institut f\"ur Optik und Quantenelektronik, Friedrich-Schiller-Universit\"at Jena, Max-Wien-Platz 1, 07743 Jena, Germany}
	\affiliation{Helmholtz-Institut Jena, Fr\"obelstieg 3, 07743 Jena, Germany}
	\author{S. Villalba-Ch\'{a}vez}
	\affiliation{Institut f\"ur Theoretische Physik I, Heinrich Heine Universit\"at D\"usseldorf, Universit\"atsstr. 1, 40225 D\"usseldorf, Germany}
	\author{A. B. Voitkiv}
	\affiliation{Institut f\"ur Theoretische Physik I, Heinrich Heine Universit\"at D\"usseldorf, Universit\"atsstr. 1, 40225 D\"usseldorf, Germany}
	\author{M. Zepf}
	\affiliation{Institut f\"ur Optik und Quantenelektronik, Friedrich-Schiller-Universit\"at Jena, Max-Wien-Platz 1, 07743 Jena, Germany}
	\affiliation{Helmholtz-Institut Jena, Fr\"obelstieg 3, 07743 Jena, Germany}
	\author{C. M\"uller}
	\affiliation{Institut f\"ur Theoretische Physik I, Heinrich Heine Universit\"at D\"usseldorf, Universit\"atsstr. 1, 40225 D\"usseldorf, Germany}
	\date{\today}
	\begin{abstract}
		Near-future experiments intend to detect strong-field Breit-Wheeler pair creation from the collision between bremsstrahlung bursts containing GeV-$\gamma$ quanta and high-intensity laser pulses. In this theoretical study, we investigate the influence of laser focusing, radiation reaction and a broad bremsstrahlung $\gamma$ spectrum on the energy and angular distributions of created pairs. Understanding the role of these inherent reaction attributes provides relevant insights for experimental detection strategies and data interpretation. We show that the inclusion of radiation reaction leads to a narrow energy spectrum of the yielded particles, whose maximum is shifted to substantially lower energies as compared to the case in which radiative energy losses are ignored. The broad bremsstrahlung spectrum also has distinct influence on the particle distributions, whereas the impact of laser focusing turns out to be rather moderate in the considered parameter regime. 
	\end{abstract}
	
	\maketitle
	\section{Introduction}
	
	The possibility of materializing quantum vacuum fluctuations into an electron-positron pair through the collision of two photons \cite{BW} — a phenomenon known in the literature as linear Breit-Wheeler process — is among the predictions of quantum electrodynamics (QED) that contributed to replacing the classical inert vacuum perception by the modern picture resembling a polarizable, absorptive dielectric medium. This view was further extended by subsequent investigations. Noteworthy are those related to the pair creation process in the field of a monochromatic plane wave background \cite{Reiss,Nikishov}, which put forward reaction channels $\omega^\prime + n\omega \to e^++e^-$ in which multiple strong field photons can be absorbed, thus giving compelling theoretical evidence that the quantum vacuum mediates nonlinear electromagnetic interactions at the most fundamental level.

	In the context of Breit-Wheeler pair creation, the nonlinear feature is predicted to manifest in both the perturbative weak field ($\xi\ll 1$) and the nonperturbative strong-field ($\xi\gg 1$) regimes. This classification refers to the coupling with the laser field, which is measured by the laser intensity parameter $\xi = |e\mathcal{E}_0|/(m\omega)$. Here, $\mathcal{E}_0$ is the laser peak field strength, $\omega$ the laser frequency, whereas $m$ and $e < 0$ are the electron mass and charge, respectively.\footnote{ Throughout, we use relativistic units  with  $c= \hbar =4\pi\epsilon_0= 1$ and denote four-products with Dirac $\gamma$-matrices by the Feynman slash notation.} In the perturbative case $\xi\ll 1$, the rate linked to a single nonlinear event scales as $\mathcal{R} \sim \xi^{2n}$ and is thus suppressed as the number of absorbed laser photons $n$ increases. Laser-induced pair production was experimentally realized at $\xi \approx 0.4$ by the SLAC E-144 collaboration \cite{SLAC}, observing a power-law rate scaling with $\xi^{10}$ characteristic for a few-photon process \cite{Note}.
	
	When a single pair is produced in the nonperturbative regime for $\xi\gg 1$, a very large number of laser photons is expected to be absorbed \cite{BWrecent}. The total production rate in this regime is proportional to $\mathcal{R} \propto \exp[-8/(3\kappa)]$, provided $\kappa\ll 1$ holds for the quantum nonlinearity parameter. Assuming a counterpropagating beam geometry, the latter is given by $\kappa = 2\omega^\prime  \mathcal{E}_0 /(m\mathcal{E}_c)$, with the critical field strength $\mathcal{E}_c = m^2 /|e| \approx 1.3 \times 10^{16}$ Vcm$^{-1}$. The strong exponential damping of the production rate has so far prevented an experimental observation. The difficulty arises from the lack of experimental conditions for merging at the same place a strong-field source with $\xi\gg 1$ and a high-energy photon reservoir with photon energies satisfying $\omega^\prime \gtrsim m\mathcal{E}_c /\mathcal{E}_0$.  Although the field strengths attainable from even the most powerful laser sources lie well below the QED characteristic scale $\mathcal{E}_0 \sim 10^{-3} \mathcal{E}_c$ \cite{CoReLS}, when brought into collision with a beam of high-energy $\gamma$ photons with $\omega^\prime \ge O(1)$\,GeV, the previous condition can be fulfilled, rendering the exponent in $\mathcal{R}$ of the order of one and thus enabling an observation of the strong-field Breit-Wheeler pair creation.
	
 	Corresponding experiments are currently under way in various high-field laboratories worldwide, including the Extreme Light Infrastructure \cite{ELI}, the E-320 project at SLAC \cite{E320}, the Rutherford Appleton Laboratory \cite{RAL}, the second phase of the LUXE project at DESY \cite{LUXE}, and the Center for Advanced Laser Applications (CALA) \cite{CALA}. The latter two collaborations plan to utilize GeV $\gamma$-quanta generated through bremsstrahlung. 
	
	Motivated by these experimental prospects, nonlinear, nonperturbative Breit-Wheeler pair production in collisions of high-intensity laser pulses with bremsstrahlung $\gamma$-photons has been theoretically studied in recent years, this way further elaborating the original proposal \cite{Reiss1971}. Total pair yields have been calculated for $\xi\sim 1$ \cite{Hartin, Eckey} and $\xi\gg 1$ \cite{Blackburn, Golub}, in particular in view of the upcoming  experiments at DESY and CALA. For the first phase of the LUXE experiment at $\xi\lesssim 1$, also energy spectra and angular distributions of the produced particles have been obtained \cite{Eckey}. In contrast to the traditional  treatments of nonlinear Breit-Wheeler pair production by monoenergetic $\gamma$-photons and a plane laser wave \cite{Reiss,Nikishov}, these recent studies take the broad energy spectrum of bremsstrahlung into consideration.  For $\xi\gg 1$, it is also necessary to account for the laser pulse's focussing \cite{Blackburn, Golub, PiazzaBW, Riconda}, which is experimentally  unavoidable as high peak intensities are required, making the strong field description deviate considerably from a plane-wave model.
	
	In the present paper, we extend the previous studies of strong-field [$\xi\gg1$] Breit-Wheeler pair production in laser-bremsstrahlung collisions by investigating the energy and angular distributions of the created particles.	 
	This kind of information is relevant for the design of upcoming experiments. Apart from the impacts of tight laser focusing and broad bremsstrahlung spectrum, there is another important aspect to be considered: The particles are created with high energy ($\sim\mbox{GeV}$) and, directly after their creation, still subject to the high-intensity laser pulse ($\xi\sim 10^2$, $\kappa\sim 1$). Their dynamics is therefore strongly affected by radiation reaction forces exerted by the laser field \cite{LL3, Koga2005, LLPiazza, Lehmann, Ruhl2014, Gonoskov, Hosaka,PiazzaTS}, whose relevance can be estimated in the present scenario by the parameter $R\approx \frac{1}{4}\alpha\kappa\xi\sim 1$, with the fine-structure constant $\alpha=e^2$. These effects have to be taken into account in order to predict the particles' spectra, as can be measured at a detector. The planned experiments, e.g. at CALA, on nonperturbative Breit-Wheeler pair production will therefore inherently probe, as a byproduct, radiation reaction effects as well. In connection, we note that such effects have recently been observed in dedicated studies by colliding a beam of relativistic electrons with high-intensity laser pulses \cite{Poder, Cole2018}.

	Our paper is organized as follows. We first briefly review the nonlinear Breit-Wheeler process, caused by a $\gamma$-photon colliding with a plane laser wave. Then, in Sec.~\ref{PPPFLPSec},  we describe Di Piazza's treatment in Ref.~\cite{PiazzaBW}, which incorporates a focused pulse as the strong-field background. The effect of bremsstrahlung $\gamma$-photons' broad frequency spectrum on the pair production process is covered in Sec.~\ref{BremsstrahlungSec}.  Afterwards, in Sec.~\ref{RRSEC}, we put forward how the radiation reaction effects of the produced particles can be taken into account within the resulting energy spectrum. Section~\ref{resultdddd} is devoted to the results. We first examine, in Sec.~\ref{WRR}, the energy distribution resulting from the collision between bremsstrahlung  quanta and a high-intensity laser pulse, taking into consideration the effect due to the laser's focusing. The impact of the radiation reaction on the spectrum is then discussed in Sec.~\ref{SWRRE}. Next, in Sec.~\ref{ADAFE}, the angular distribution of the created pairs is investigated. Finally, we give our conclusions and an outlook on future research in Sec.~\ref{conclusion}.
	
	\section{Theoretical considerations \label{secII}}
	
	The S-matrix element for the strong-field Breit-Wheeler process with a high-energy $\gamma$-photon of frequency $\omega^\prime$, wave vector $k^{\prime\mu}=(\omega^\prime,\textbf{k}^\prime)$ and polarization vector $\epsilon^\mu_{k^\prime,l}$ is given by
	\begin{eqnarray}
		S_{\text{fi}} = -ie\sqrt{\frac{4\pi}{2\omega^{\prime}}}\int \text{d}^4x\;\text{e}^{-ik^{\prime} \cdot x} \bar{\Psi}_{ p,\mathrm{s}} \slashed{\epsilon}_{k^\prime,l} \Psi_{-p^\prime,-\mathrm{s}^\prime}.\label{eqq1}
	\end{eqnarray}
	Here, $p^\mu=(\varepsilon, \textbf{p})$ denotes the the electron four momentum, whereas $p^{\prime\mu}=(\varepsilon^\prime, \textbf{p}^\prime)$ is  the positron four momentum. Using this, the number of pairs  per $\gamma$-photon is obtained as
	\begin{eqnarray}
		&&\text{d}N(\omega^\prime)= \frac{1}{2}\sum_{l,s,s^\prime}|S_{\text{fi}}|^2 \frac{\text{d}^3p}{(2\pi)^3}\frac{\text{d}^3p^\prime}{(2\pi)^3},
	\end{eqnarray}
	where we have averaged over the photon polarization and have summed over the lepton spins $s$ and $s^\prime$. The S-matrix in Eq.~\eqref{eqq1} is formulated in the Furry picture, where the lepton states $\Psi_{p,s}$ and $\Psi_{-p^\prime,-\text{s}^\prime}$ are dressed by the laser field. If the latter is of plane-wave shape, the laser-dressed states are given exactly by the well-known Volkov states \cite{LL}, which are quasiclassical wave functions \cite{PiazzaWKB} of the form $\Psi_{\pm p}\propto \exp(iS_\pm)$, with the classical action $S_\pm$ in the laser field. The coupling of the dressed lepton states to the $\gamma$-photon field is described in the first order of perturbation theory. The latter is applicable when $\alpha\kappa^{2/3} < 1$ holds \cite{Narozhny}, which is fulfilled for the parameters under consideration.
	
	\subsection[A.]{Pair creation in a tightly  focused laser field \label{PPPFLPSec}}
	
	To calculate the S-matrix element of the Breit-Wheeler process in a focused laser field, one can consider generalized Volkov states which are obtained by the quasiclassical WKB method. Following the treatment developed in Ref.~\cite{PiazzaBW}, we shall assume that the high-energy $\gamma$-photon and the laser pulse counter-propagate, and describe the latter by the four-potential 
	\begin{eqnarray}
		\mathcal{A}_\perp(\textbf{x}) = \int_{T}^{\infty} d\tilde{T} \left[ \textbf{E}_\perp(\tilde{T},\textbf{x}_\perp) + \textbf{e}_z \times \textbf{B}_\perp(\tilde{T},\textbf{x}_\perp)   \right]
	\end{eqnarray}
	in lightcone coordinates $\textbf{x} = (T, \textbf{x}_\perp)$. In our calculation the laser pulse propagates in the negative $z$ direction, so that $T= (t+z)/2$ and $\textbf{x}_\perp = (x,y)$. 
	The electric field is taken in paraxial approximation as \cite{Salamin}
	\begin{equation}\label{focusfield}
		\begin{split}
			&\textbf{E}_\perp =\mathcal{E}_0 \textbf{e}_x \text{e}^{-\frac{r^2}{w^2(z)}} \frac{\text{e}^{-\left(\sqrt{2\ln(2)}\frac{\Phi}{\omega\tau}\right)^2}}{\sqrt{1+\zeta(z)^2}}\\
			&\qquad\quad\times \sin\left(\Phi -\zeta(z)\frac{r^2}{w^2(z)}+\arctan\left(\zeta(z)\right)\right). 
		\end{split}
	\end{equation}
	The nontrivial magnetic field component linked to  this field configuration shares the magnitude of  the electric field [$B_y = E_x$]. In the expression above, $\Phi = \omega (t+z)$ is the laser phase,  $\mathcal{E}_0$ refers to the field amplitude, $z$  is the longitudinal component, and  $r=\sqrt{x^2+y^2}$ denotes the transversal component. The parameter $w_0$ gives the beam waist size at the focal point $z=0$. Due to the factor $\zeta(z) =z/z_R$, where $z_R = \pi w_0^2/\lambda$ denotes the Rayleigh length, the general beam width $w(z) = w_0 \sqrt{1+\zeta(z)^2}$ depends on the longitudinal component. When the $z$ dependence that comes together with $z_R$ is ignored,  a field with solely transversal focussing $\sim {\rm e}^{-r^2/w_0^2}$ results (infinite Rayleigh-length approximation). For finite Rayleigh length, the field features in addition longitudinal focussing. Besides,  the temporal extension of the pulse $\tau$ is taken at FWHM from the intensity.
	
	Along the lines of Ref.~\cite{PiazzaBW}, which uses the condition $\omega^\prime \gg m\xi > m$, we obtain the  angular distribution
	\begin{equation}\label{angulardistribution}
		\begin{split}
			&\frac{d^2N(\omega^\prime)}{d\vartheta d\varphi} = \frac{\rho_\gamma \alpha}{\sqrt{3}\pi {\omega^\prime}^2\omega} \int d\varepsilon \int d\Phi \int dr\;\varepsilon^2 r \sin(\vartheta)\\
			& \qquad\qquad \times  f(\textbf{x}) b(\textbf{x}) \left[1 + \frac{ \varepsilon^2 + {\varepsilon^\prime}^2}{\varepsilon\varepsilon^\prime}  f^2(\textbf{x}) \right]\\
			& \qquad\qquad\times \text{K}_{\nicefrac{1}{3}}\left(\frac{2}{3} b(\textbf{x}) f^3(\textbf{x})\right)
		\end{split}
	\end{equation}
	and the energy spectrum
	\begin{equation}\label{energydistribution}
		\begin{split}
			\frac{dN(\omega^\prime)}{d\varepsilon} &= \frac{\rho_\gamma \alpha}{\sqrt{3} {\omega^\prime}^2\omega} \int d\Phi \int dr~ r \bigg[\frac{\varepsilon^2 + {\varepsilon^\prime}^2}{\varepsilon\varepsilon^\prime}  \\
			&\times \text{K}_{2/3}\left(\frac{2}{3} b(\textbf{x})\right)+ \int_{\frac{2}{3}b(\textbf{x})}^{\infty}dv~ \text{K}_{\nicefrac{1}{3}}(v)  \bigg]
		\end{split}
	\end{equation}Here, $\text{K}_{\nicefrac{1}{3}}(x)$ and $\text{K}_{\nicefrac{2}{3}}(x)$ are  modified Bessel functions \cite{NIST}, whereas $f(\textbf{x})= \sqrt{1+ [\textbf{p}_\perp^\prime + e \mathcal{A}_\perp]^2/m^2}$, and $b(\textbf{x})={\omega^\prime}^2/(\varepsilon\varepsilon^\prime \kappa(\mathbf{x}))$ with $\kappa(\mathbf{x})=\frac{\omega^\prime}{m\mathcal{E}_c}\left|\frac{\partial \mathcal{A}_\perp}{\partial T}\right|$ referring to the local value of the quantum nonlinearity parameter.  In the formulas above $\rho_\gamma$ stands for the density of incoming $\gamma$ photons, and the angle $\vartheta$ should be understood relative to its direction of propagation. It should be mentioned that, in line with Ref.~\cite{PiazzaFocus}, the approximation $z\approx t \approx T$ was made for the use of Eq.~\eqref{focusfield}.
	
	We note that the limit  $w_0 \rightarrow \infty$ in Eq.~\eqref{focusfield} makes the field background  a  plane wave pulse
	\begin{eqnarray}\label{PWBack}
		\textbf{E}_\perp(\Phi) &=& \mathcal{E}_0 \textbf{e}_x  \text{e}^{-\left(\sqrt{2\ln(2)}\frac{\Phi}{\omega\tau}\right)^2} \sin\left(\Phi\right).
	\end{eqnarray}
	This strong field model will be used in the following for assessing differences from the outcomes associated with a laser pulse that is tightly focused [see Eq.~\eqref{focusfield}].  For this, we shall suppose that both types of backgrounds carry the same energy, which is the case provided an effective transversal area $\Sigma_{\mathrm{eff}}= \pi w_0^2/2$ for the plane wave is introduced \cite{Golub}.
	
	\subsection[C.]{Inclusion of bremsstrahlung\label{BremsstrahlungSec}}
	
	In the forthcoming analysis, we shall suppose that bremsstrahlung gamma quanta from the interaction between a monoenergetic few-GeV electron beam and a high Z-target, whose thickness is much smaller than the characteristic radiation length $L_\text{rad}$ of the material, are firstly generated. This is primarily motivated by the ongoing setup at CALA, which aims to produce electron-positron pairs from the collision between  this highly energetic photon source and a tightly focused laser pulse. 
	\begin{figure}[h]
		\centering
		\includegraphics[width=0.50\textwidth]{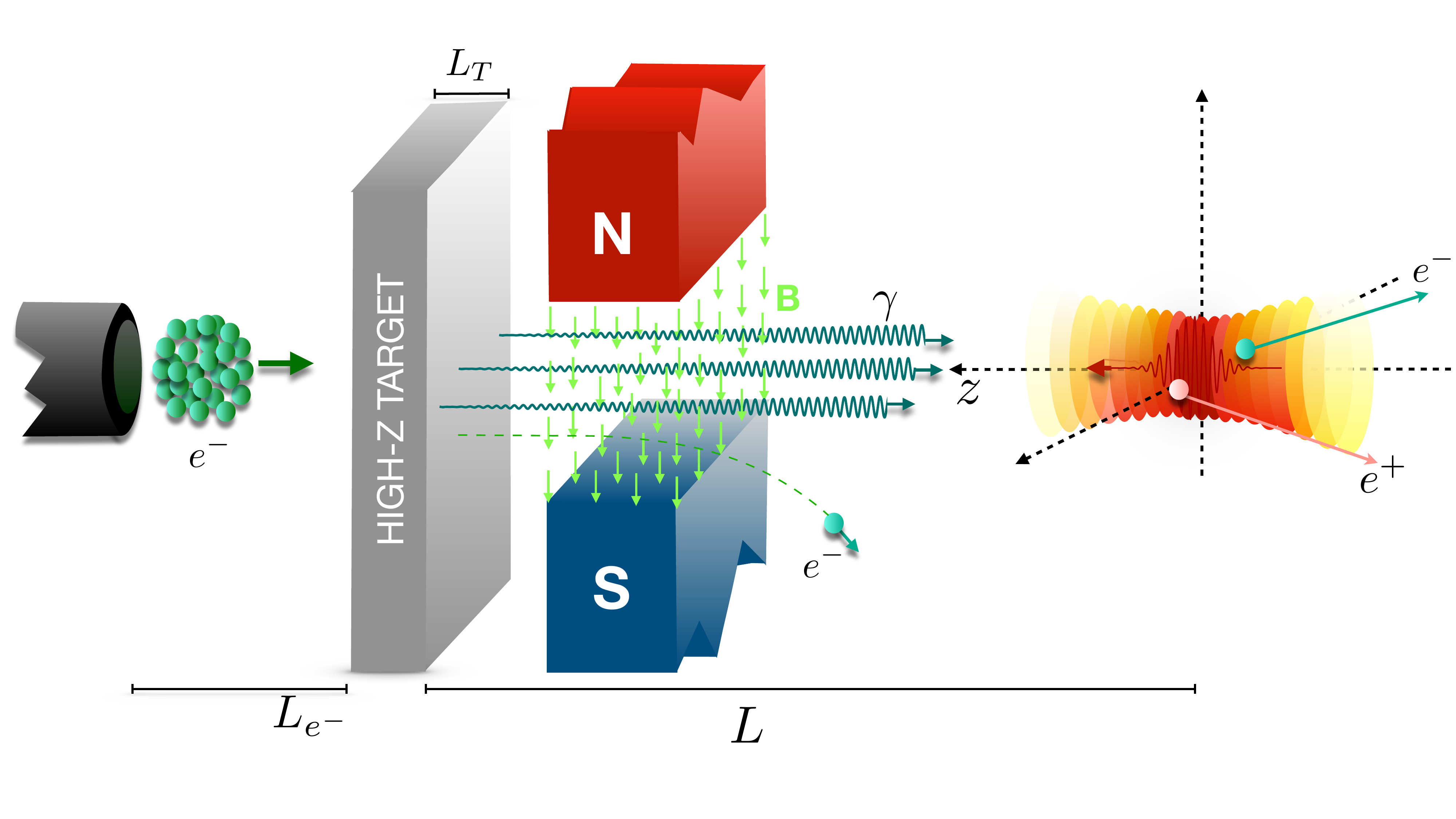}
		\caption{\label{Fig:1}
			Experimental outline for the strong-field nonlinear Breit-Wheeler pair production process from the collision of bremsstrahlung gamma photons and a high-intensity laser pulse. The bremsstrahlung burst used in the setup will be produced as a consequence of the interaction between a monoenergetic electron beam and a thin high-$Z$ target. A magnet behind the target will be used to deflect incoming electrons away from  the region where the photon-photon collision takes place \cite{CALA}.}
	\end{figure}
	Fig.~\ref{Fig:1} shows a sketch of the corresponding experimental configuration.  In the  proposed scenario the  electron beam is expected to undergo almost no spatial divergence along the interaction with the target owing to both the ultra-relativistic nature of its constituents ($E_0=2.5$\,GeV) and the fact that its extension would be comparable to the target thickness.  It is therefore  reasonable to assume that the emission of bremsstrahlung photons occurs inside the electron beam volume, within a spreading angle determined, to a large extent, by the inverse electron Lorentz factor $\Theta_\gamma \approx 1/\gamma_0 \propto O(1)~$mrad.  As a result, the vast majority of bremsstrahlung photons would be emitted in the direction of propagation of the incident electron beam and a head-on collision with the laser pulse, as required by the formulation outlined in Sec.~\ref{PPPFLPSec}, can be assumed.
	
	In the present scenario, the angular and energy distributions associated with the produced electron-positron pair  are obtained by integrating  Eqs.~(\ref{angulardistribution}) and  (\ref{energydistribution}) over the  photon energy $\omega^\prime$, weighted by the Bremsstrahlung spectrum $d\mathcal{N}_\gamma/d\omega^{\prime}$ of a radiating electron of incident energy $E_0$ calculated within the complete screening approximation  \cite{Tsai}. Consequently, 
	\begin{equation}\label{Distributionsplusbremms}
		\begin{split}
			\frac{d^2N}{d\vartheta d\varphi}&=\int_{\mathfrak{f}_{\text{min}}}^{1}\text{d}\mathfrak{f}\;I_\gamma(\mathfrak{f},\ell) \left.\frac{d^2N(\omega^\prime)}{d\vartheta d\varphi}\right\vert_{\omega^\prime=\mathfrak{f}E_0},\\
			\frac{dN}{d\varepsilon} &=\int_{\mathfrak{f}_{\text{min}}}^{1}\text{d}\mathfrak{f}\;I_\gamma(\mathfrak{f},\ell)\left.\frac{dN(\omega^\prime)}{d\varepsilon}\right\vert_{\omega^\prime=\mathfrak{f}E_0}, 
		\end{split}
	\end{equation}
	where we have gone over to the normalized energy variable $\mathfrak{f}=\omega^\prime/E_0$. From now on, we will use the thin target approximation, in which case
	\begin{eqnarray}
		I_\gamma(\mathfrak{f},\ell)=\frac{d\mathcal{N}_\gamma}{d\mathfrak{f}}\approx \frac{\ell}{\mathfrak{f}}\left( \frac{4}{3} - \frac{4\mathfrak{f}}{3} +\mathfrak{f}^2 \right).
	\end{eqnarray}
	The expression above applies for a normalized target thickness $\ell=L_T/L_{\text{rad}} < 1$. Due to the energy conservation in the process, the minimum value of $\mathfrak{f}$ must be  $\mathfrak{f}_{\text{min}}= \varepsilon/E_0$ for a given energy $\varepsilon$ of the created electron. Note in this regard that, for the parameters used in Sec.~\ref{resultdddd}, the energy of the pair is essentially given by the $\gamma$ photon energy, i.e. $\omega^\prime \approx \varepsilon + \varepsilon^\prime$.
	
	We further note that, in our parameter regime, the main contribution to the pair creation stems from the region close to $\mathfrak{f}\approx 1$ (see Ref.~\cite{Golub}). Therefore, the condition $\omega^\prime\gg m\xi$ is safely met in Eq.~\eqref{Distributionsplusbremms} for $\xi \lesssim 10^2$. While the broad spectrum of bremsstrahlung contains apart from high-energy photons also a large fraction of low-energy photons, the latter effectively do not lead to pair production. 
	
	\subsection[D.]{Inclusion of radiation reaction \label{RRSEC}}
	
	The previous description neglects the effect of radiation reaction on the created particles because the classical action $S_\pm$ in the Volkov states only accounts for the interaction between the electron (or positron) and the external field in terms of the Lorentz equation of motion. However, directly after their creation, the electron and positron are still exposed to the very strong electromagnetic fields of the laser pulse. In the parameter regime of interest, their classical dynamics is not properly described by the Lorentz equation. Because of the strong forces exerted by the laser field, the particles experience violent acceleration, which causes energy losses by emission of radiation. These radiation reaction effects on the particle dynamics have to be taken into account, in order to obtain reliable predictions for the momentum distributions that can be measured on a detector (which is placed far outside the region where the  $\gamma$-quanta and the focused laser interact). 
	
	We shall  address the analytically solvable radiation reaction problem in a plane-wave background field \cite{LLPiazza}. Depending on the electron quantum nonlinearity parameter $\chi_e=(1-\cos\theta_e)\gamma_e\mathcal{E}_0/\mathcal{E}_c$, with $\gamma_e = \varepsilon/m$ and $\theta_e$ giving the angle between the laser propagation direction and the electron momentum, the radiation reaction must be treated appropriately. For $\chi_e\ll 1$, the electron's motion---interacting with an external background and the field it radiates---can be described by the Landau-Lifshitz equation (LL), which is given by \cite{LL3}:
	\begin{eqnarray}
		m\frac{du^\mu}{ds} &=& e F^{\mu\nu}(x)u_\nu + R^\mu,\label{Lorentzeqa}
	\end{eqnarray} 
	with the radiation reaction force
	\begin{equation}
		\begin{split}
			&R^\mu =  \frac{2}{3} \alpha \bigg[ \frac{e}{m} (\partial_\delta F^{\mu\nu})u^\delta u_\nu  \\
			&\qquad\quad  -\frac{e^2}{m^2} F^{\mu\nu}F_{\delta\nu} u^\delta +\frac{e^2}{m^2}(F^{\delta\nu} u_\nu)(F_{\delta\lambda} u^\lambda)u^\mu \bigg]. 
		\end{split}
	\end{equation}
	Here, $u$ denotes the four-velocity of the electron and $s$ its proper time. Moreover $F_{ \mu\nu}$ is the electromagnetic field tensor linked to the strong field background.
	
	If $\chi_e$ is not small, which is the case in our parameter regime ($\chi_{e} \approx 1$ for $\varepsilon\approx 1$\,GeV), the loss of energy of the electric charge is overestimated by Eq.~\eqref{Lorentzeqa}, as there is no classical upper limit for the energy of the continuously emitted radiation. However, in quantum terms, the energy of an emitted photon is bounded by the kinetic energy of the electron. This results in a sharp quantum cutoff in the emitted radiation spectrum, reducing the amount of emitted energy. To take this cutoff into account $R^\mu$ is multiplied by a weighting function $g(\chi_e)$, called Gaunt factor $g(\chi_e) =I_Q/I_C$, which is defined  by the ratio between  quantum radiation intensity  in  a constant crossed field  $I_Q$ and  the classical radiation intensity $I_C$. Approximately one finds \cite{Poder}
	\begin{equation}
		g(\chi_e)  \approx \left[1 + 4.8(1+\chi_e)\ln(1+1.7\chi_e) + 2.44\chi_e^2 \right]^{-\frac{2}{3}}. 
	\end{equation}	
	
	This Gaunt factor makes a further analytical solution of the modified LL equation difficult and a numerical consideration is generally necessary. However, one way to proceed analytically is to assume that $\gamma_e$ changes only slowly compared to the phase of the laser field, which implies $g\left(\chi_e(\Phi)\right)$, where $\chi_e(\Phi)$ takes the field variation into account. This enables us to follow the steps outlined in Ref.~\cite{LLPiazza} and, as a result, approximate the LL equation's solution using the one found on a plane-wave background:
	\begin{equation}
		\begin{split}
			&u^\mu(\Phi) =\frac{1}{h(\Phi)}\bigg[u_0^\mu + \frac{1}{2\rho_0}[h^2(\Phi)-1]n^\mu  \\
			& \qquad\qquad\quad+ \frac{1}{\rho_0} I(\Phi)\frac{e f^{\mu\nu}}{m}u_{0,\nu}+ \frac{1}{2\rho_0}\xi^2 I^2(\Phi)n^\mu\bigg].
		\end{split}\label{uphi}
	\end{equation}
	In the expression above, $u_0$ denotes the initial velocity, $\rho_0=n_\mu u_0^\mu$, $n^\mu=k^\mu/\omega$ the four-wave vector of the strong pulse, the four-potential of which $A_\mu(\Phi)=a_\mu \psi(\Phi)$ is modulated by the function $\psi(\Phi)$, with $a_\mu$ referring to its amplitude. In this context, the electromagnetic tensor linked to the strong field  $F_{\mu\nu}(\Phi)=f_{\mu\nu}\psi^\prime(\Phi)\omega$ with $f_{\mu\nu}=n_\mu a_\nu-n_\nu a_\mu$ and $\psi^\prime(\Phi)= -\exp\left[-\left(\sqrt{2\ln(2)}\Phi/(\omega\tau)\right)^2\right] \sin\left(\Phi\right)$ (see  Eq.~\eqref{PWBack}). The expression above depends on the functions
	\begin{equation}
		h(\Phi)= 1 +\frac{2}{3}\frac{\rho_0\omega}{ m}\alpha\xi^2 \int_{\Phi_i}^{\Phi} d\tilde{\Phi}~ g(\chi_e(\tilde{\Phi}))\psi^2(\tilde{\Phi}) 
	\end{equation}
	and
	\begin{equation}
		\begin{split}
			&I(\Phi) = \int_{\Phi_i}^{\Phi} d\tilde{\Phi}\Big[h(\tilde{\Phi})\psi^\prime(\tilde{\Phi})\\
			&\qquad\qquad\qquad\qquad+\frac{2}{3}\frac{\rho_0\omega}{m}\alpha g(\chi_e(\tilde{\Phi}))\psi^{\prime\prime}(\tilde{\Phi}) \Big].
		\end{split}
	\end{equation}
	If we set $g(\chi_e)=1$, we arrive at the same solution as in \cite{LLPiazza}. The upper integration limit is chosen as the one at the end of the pulse i.e., the limit $\Phi\to\infty$ is taken at the end of the calculations. 
	
	\begin{figure}[h]
		\centering
		\includegraphics[width=1\linewidth]{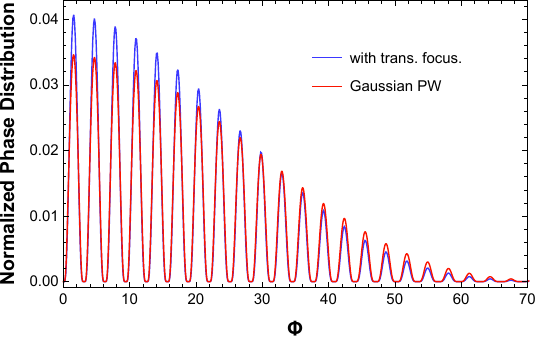}
		\caption{\label{Fig2}Normalized phase distribution $\frac{1}{N}\frac{dN}{d\Phi}$ of the Breit-Wheeler pair production process from the collision of a single gamma quantum with energy $\omega^\prime=E_0=2.5\;\rm GeV$ and a strong laser field.  While the curve colored in blue is linked to the case in which the background is  a focused pulse, the one in red corresponds to   the situation in which it is described by a Gaussian plane wave. The benchmark parameters in Tab.~\ref{table} have been applied in order to produce this picture.
		}
	\end{figure}
	
	The unspecified parameter $\Phi_i$ gives the phase point from which onwards the radiation reaction has to be taken into account. Approximated values for it can be found by investigating the behavior of the normalized phase distribution. This dependence is shown in Fig.~\ref{Fig2} for the cases in which the strong field background is either considered a Gaussian plane wave (red curve) or a focused laser pulse (blue curve). Within the phase interval $\Delta\Phi_n=[n\pi,(n+1)\pi]$, both results are typified by prominent peak formations near $\Phi_n=(2n+1)\pi/2$. As at these locations, the probability of producing electron-positron pairs is higher, they represent an appropriate estimate for the parameter $\Phi_i$.  We note that the use of this choice in Eq.~\eqref{uphi} is consistent with the intuitive picture where pairs formed early in the pulse are exposed to radiation reactions for a longer period of time than those created at the ending pulse tail.   It is also  worth noting that the peaks associated with each strong field model have similar heights at $\Phi_n=(2n+1)\pi/2$. This indicates that each phase interval contributes nearly the same amount to the total number of produced pairs in both strong field models.  Therefore, the outcomes linked to the radiation reaction impact in the situation where the background is a Gaussian plane wave may be considered as close approximation to the realistic scenario driven by a tightly focused laser pulse. It is worth mentioning that the initial parameters $u^\mu_0$ can be determined from the spectra per phase interval obtained by Eq.~\eqref{angulardistribution} by taking into account Eq.~\eqref{PWBack}.
	
	The outlined procedure enables us to incorporate the radiation reaction into the particles' momentum distributions. First, for each phase interval, the momenta (inside the field) of the particles created at phase $\Phi_n$ are determined. Then these initial values are propagated through the remaining laser pulse with the help of Eq.~\eqref{uphi}. By summing over all phase intervals, weighted by the respective production probabilities, the momentum distributions outside the field, including the radiation reaction, are obtained.
	
	We note that the semiclassical approach to radiation reaction based on the LL equation combined with the gaunt factor provides good agreement with the experimental energy spectra obtained in Ref.~\cite{Poder}. Collisions of electrons with a maximum initial energy of $\approx 2$\,GeV and an intense laser pulse with $\xi \approx 10$ were studied there. While including the electron recoil effect, the semiclassical model does, however, not account for the probabilistic nature of the photon emissions. To include such  stochasticity effects, a fully quantum description of the radiation reaction would be required.

	\section{Results and Discussion\label{resultdddd}}
	
	This section  is devoted to analyze the results obtained for the  energy spectra  and the angular distributions of the created particles. We start from the two cases of a plane laser wave with either bremsstrahlung $\gamma$-rays or with a single $\gamma$-photon as references, in order to reveal via comparisons, on the one hand, the impact of laser focusing effects and, on the other hand, the influence of radiation reaction. With respect to laser focusing, we consider both the transversal as well as the longitudinal focusing. We point out that most of the curves in the following figures have been normalized to a height of unity to facilitate their comparison.
	
	For the analysis we use the values given in Tab.~\ref{table}, unless otherwise stated. These benchmark parameters fulfill the condition $\omega^\prime \gg m\xi \gg m$.
	\begin{table}[h]
		\centering
		\begin{tabular}{ |p{6cm}|p{1.5cm}|}
			\hline
			\text{Incident electron energy $E_0$} & 2.5~GeV \\
			\hline
			\text{Normalized target thickness $\ell$} & 0.015 \\
			\hline
			\text{Distance travelled by bremsstrahlung $L$} & 0.5~m \\
			\hline
			\text{Wavelength of the strong pulse $\lambda$} & 0.8~$\mu$m \\
			\hline
			\text{Pulse waist size $w_0$} & 2~$\mu$m \\
			\hline
			\text{Pulse length $\tau$} & 30~fs \\
			\hline
			\text{Laser intensity parameter $\xi$} & 70\\
			\hline
		\end{tabular}
		\caption{The parameters for the experiment to be performed at CALA, applying a linearly polarized, tightly focussed laser pulse (see Ref.~\cite{CALA}). These values are assumed in our numerical calculations. \label{table}}
	\end{table}
	
	\subsection{Energy spectra without radiation reaction\label{WRR}}
	
		\begin{figure}[h]
		\centering
		\includegraphics[width=1\linewidth]{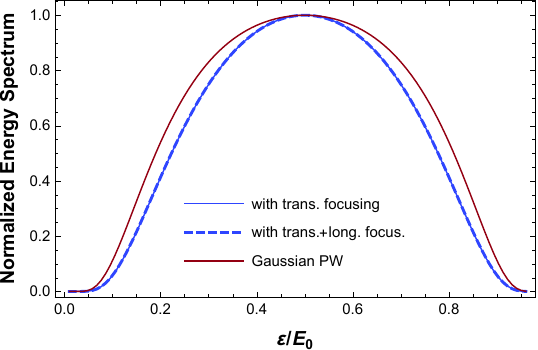}
		\caption{\label{fig3}The normalized energy spectrum for a single $\gamma$ photon with $\omega^\prime = 2.5\, \text{GeV} =E_0$ is shown for the focused laser field with transversal focusing as a solid blue curve and with transversal and longitudinal focusing as a dashed blue curve. The spectrum for a laser field as a pulsed plane wave is shown as a solid red curve.  }
	\end{figure}
	
	In Fig.~\ref{fig3}, the dependence of the normalized energy spectrum on the scaled electron energy $\varepsilon/E_0$ is shown. This picture includes the results obtained from the collision of a single gamma quantum and a strong laser background, for which the following models were adopted: a plane wave in red, a focused pulse with solely transverse focusing (solid blue), and a pulse with both transverse and longitudinal focusing (dashed blue curve). The maximum of all curves is located at $\varepsilon/E_0 \approx 0.5$, which is due to the fact that the absorbed energy mainly stems from the $\gamma$-photon, implying $\omega^\prime \approx \varepsilon + \varepsilon^\prime$,  and that this energy is distributed approximately equally between the electron and positron. Both curves linked to the focused field models lie on top of each other in the normalized case, which indicates, that accounting for the transversal focusing is sufficient to predict the energy spectrum. We have checked numerically, that the number of created pairs in the case without longitudinal focusing is about $\approx 1\%$ higher than in the  scenario where it is included. Compared to the focused laser field models, the spectrum resulting for a plane-wave pulse (red curve) is significantly broader. Therefore, the spectrum of the particles that are produced is narrowed as a result of the laser focussing (see also \cite{PiazzaBW}). This phenomenon can be understood by looking at the field intensity. While for the pulsed plane wave the intensity is kept constantly high within the whole transverse interaction plane, the intensity for a focused Gaussian profile changes from its maximum at the center to the minimal values at the edges. 
	
\begin{figure}[h]
		\centering
		\includegraphics[width=1\linewidth]{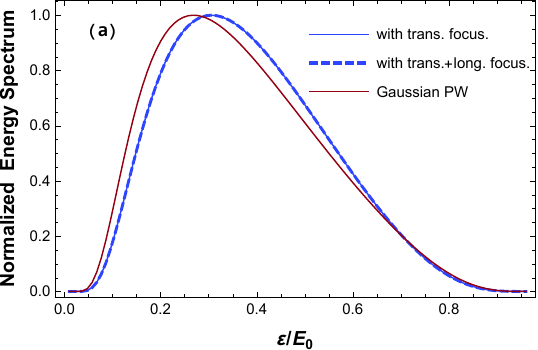}
		\qquad
		\includegraphics[width=1\linewidth]{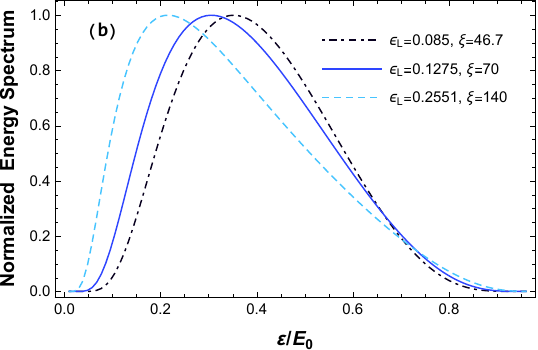}
		\caption{The normalized energy spectrum including bremsstrahlung is shown for the focused laser field without longitudinal focusing as a solid blue curve. In panel (a), comparisons with an also longitudinally focussed laser pulse and with a plane-wave laser pulse are shown. The intensity parameter is choosen as $\xi=70$ for all curves. The same color coding and normalization as in Fig.~3 applies. In panel (b) the normalized energy spectrum for the focused laser field without longitudinal focusing is shown for different pulse waist sizes $w_0$, where the pulse energy $W_G\propto \xi^2 w_0^2$ is kept constant. The resulting different diffraction angles and intensity parameters are indicated in the legend.\label{Fig4}}
	\end{figure}

	We now consider the energy spectrum taking into account the bremsstrahlung spectrum, which is shown in Fig.~\ref{Fig4}. In all cases we see clear differences compared to Fig.~\ref{fig3}. While the total	energy of a pair satisfies again the relation $\omega^\prime \approx \varepsilon + \varepsilon^\prime$, the maxima of the spectrum are not located at half the value of $E_0$ anymore, but somewhat below. This is because, while $\gamma$ frequencies close to $E_0$ are most relevant \cite{Golub}, also less energetic quanta with frequencies $\omega^\prime>0.2 E_0$ give sizeable contributions to the pair yield. In Fig.~4a we see the comparison between the focused laser field and the pulsed plane wave. The solid blue curve shows the case with purely transverse focusing, and the dashed curve the case with additional longitudinal focusing. Again, practically no difference can be seen for the two laser field models in the normalized energy spectrum. 
	
	If we now compare the focused case with the plane-wave case, which is shown as a red solid curve, we see that the maximum, considering the focusing, is at higher energies. In direct comparison, the maximum for the laser field as pulsed plane wave is at $\varepsilon/E_0 \approx 0.26$ and for the focused laser wave at $\varepsilon/E_0 \approx 0.3$. The difference in the spectra is caused by the intensity gradient of the focused laser field, previously discussed. For an increasing intensity parameter $\xi$ the maximum shifts towards smaller energies. In the focused case, pair creation processes that occur outside the laser focus in regions, where the local intensity has decreased, require larger $\omega^\prime$ values (approaching $E_0$) than in the case of a plane wave to reach a sizable probability. As a result, the typical pair energy is upshifted when the laser focusing is taken into account.
	
	In Fig.~\ref{Fig4}b, we compare the normalized energy spectrum for different focusing keeping the laser pulse energy $W_G\propto \xi^2 w_0^2$ constant. For reference, we compare $w_0 =2\,\mu m$ with values $w_0 =1\,\mu m$ and $3\,\mu m$, respectively, with the diffraction angle $\epsilon_L = 2/(w_0\omega)$. For the chosen values we get $\epsilon_L = 0.1275, 0.2551$ and $0.085$. Now, in Ref. \cite{Golub}, it has been shown that for this benchmark parameters, the total number of produced pairs increases as the laser focussing becomes tighter. Our findings complement and further extend this previous study by showing that, relative to the original parameter choice, weaker laser focusing causes a distribution in the energy spectrum that moves the maximum towards higher energies. The position of the rate maximum shifts to lower electron energies with tighter focussing and correspondingly greater laser intensity parameters. With larger $\xi$, correspondingly more photons can be absorbed from the laser field (and the local value of the quantum nonlinearity parameter is larger), so a smaller $\omega^\prime$ is sufficient to produce a pair. 
	
	From our analysis we may conclude that, for the parameters of the planned experiment at CALA \cite{CALA}, the transversal laser focusing has only very minor impact on the shape of the energy spectra of the created particles. The longitudinal laser focusing exerts practically no effect at all. This indicates that the pairs are mainly produced in the innermost focal region where $z \ll z_R$.  By integrating over the (non normalized) energy spectrum we obtain the total number of produced pairs. Assuming the same collision geometry as was discussed in \cite{Golub} we obtain a total number of $\approx 4.4\times 10^{-8}$ pairs per radiating incident electron for the benchmark parameters in Tab.~\ref{table}. This result agrees well with the prediction which was made in \cite{Golub} (see Fig.~\ref{Fig7} therein), where it was calculated by using the locally constant field approximation.

	\subsection{Energy spectra with radiation reaction\label{SWRRE}}
	
	Next, we consider the energy spectrum after the radiation reaction has acted on the created particles. In light of the very moderate impact of the laser focusing, we perform our corresponding analysis within a plane-wave model for the laser pulse. This has the advantage that we can apply in our analysis the exact analytical solution to the LL equation from \cite{LLPiazza}, corrected by the Gaunt factor from Eq.~\eqref{uphi}. 
	
	\begin{figure}[h]
		\centering
		\includegraphics[width=1\linewidth]{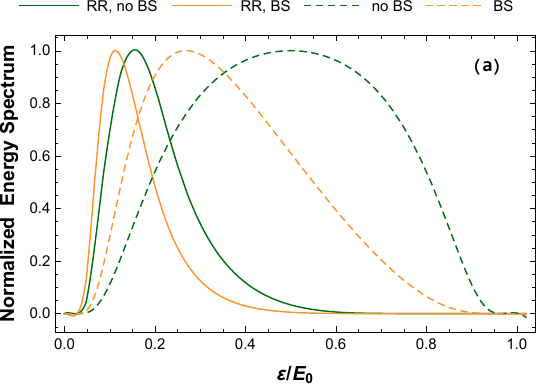}
		\qquad
		\includegraphics[width=1\linewidth]{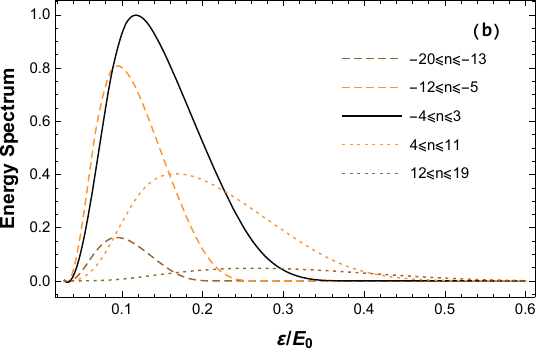}
		\caption{Normalized energy spectrum $\frac{\text{d}N}{ \text{d}\varepsilon}$  as a function of electron energy $\varepsilon/E_0$ in a pulsed plane-wave background, including radiation reaction effects. In panel (a) the solid yellow curve shows the spectrum when bremsstrahlung is taken into account and the solid green curve shows the spectrum when bremsstrahlung is not taken into account ($\omega^\prime=E_0$). For comparison the spectra without radiation reaction are given as dashed curves. In panel (b) the contributions from different laser phase intervals to the energy spectrum with bremsstrahlung, normalized to the height of the black curve are shown (compare with Fig.~\ref{Fig2}). Which phase intervals are taken into account are displayed in the legend.\label{Fig5}}
	\end{figure}	
	
	Fig.~\ref{Fig5}a shows the energy spectrum including the radiation reaction with either a single $\gamma$-photon (green curves) or when the bremsstrahlung spectrum is included (yellow curves). While the green dashed curve for the single-photon case without radiation reaction is symmetric about the point $\varepsilon = E_0/2$, the energy spectrum is shifted to substantially smaller energies and becomes asymmetric, when radiation reaction is included (green solid curve). A similar effect occurs, when bremsstrahlung is taken into account, as a comparison between the yellow dashed and yellow solid curves shows. In both cases, the shift can be explained by the fact that radiation reaction leads to energy loss, which is quite substantial in the considered scenario. Meanwhile the total number of pairs remains unchanged by inclusion of radiation reaction, which is only active in our treatment after the particles have been created.\footnote{We note that, in principle, the radiation reaction may also affect the pair creation step itself. However, this is beyond the scope of the present study.}
	
	Our approach allows us to analyze the influence of radiation reaction in dependence on the laser phase interval where a pair has been created (similarly as was done for Fig.~\ref{Fig2}). Fig.~\ref{Fig5}b shows how the energy spectrum for the case with radiation reaction and bremsstrahlung is composed of the individual contributions of the phase. For an origination of the pair in the middle phase range, shown as a black solid curve, the contribution is the largest and significantly shapes the entire spectrum. The pairs which are formed at the beginning of the pulse, shown as dashed curves, are hit most strongly by the subsequent radiation reaction and their maximum is shifted furthest to smaller energies. The further the pairs are formed  towards the end of the laser pulse, the less the radiation reaction affects them and their spectrum, which can be seen as dotted curves, becomes more and more similar to the one without radiation damping.
	
	Before moving on, it is interesting to note that the energy loss of the particles due to radiation reaction, being associated with a corresponding amount of photon emissions, has some similarity with the generation of the bremsstrahlung photons by the incident beam of electrons. While the incident electrons are scattered and decelerated upon photon emission by their interaction with the strong fields of the atomic nuclei in the converter foil, the created particles radiatively lose energy due to their interaction with the intense laser pulse. 
	
	It is also worth mentioning at this point that for pair production in laser-electron collisions, radiation reaction effects have been considered theoretically \cite{Blackburn2017,Baumann2019}. In this case, not only the created particles, but also the incident beam of electrons is subject to radiation reaction.
	
	\subsection{Angular Distributions\label{ADAFE}}
	
	We now consider the angular distribution of the created particles, referring to the angle $\vartheta$ between the electron momentum $\textbf{p}$ and the direction of the incoming $\gamma$ photon $\textbf{k}^\prime$. Since the transverse electron momentum is of order $p_\perp \sim m\xi$, while its longitudinal momentum is much larger and about $p_z \approx p_0\ \approx \omega'/2 \approx E_0/2$, one expects emission angles in the range of $\vartheta\sim m\xi/E_0 \approx 14$\,mrad.
	
		\begin{figure}[h]
		\centering
		\includegraphics[width=1\linewidth]{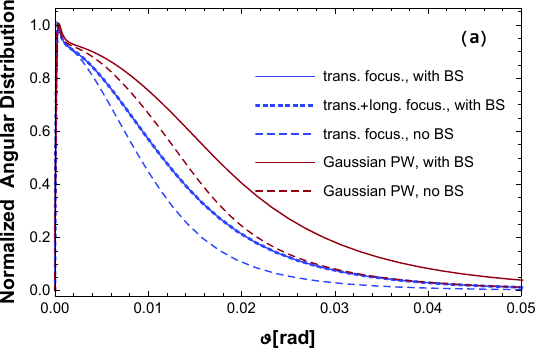}
		\qquad
		\includegraphics[width=1\linewidth]{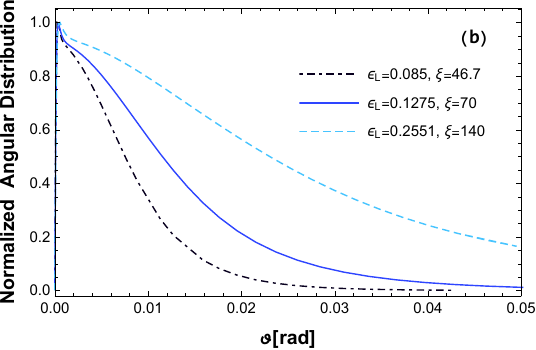}
		\caption{The normalized angular distribution including bremsstrahlung is shown for the focused laser field without longitudinal focusing as a solid blue curve. In panel (a) the comparison with a laser field as a pulsed plane wave is shown in red. The blue dotted curve displays the spectrum with longitudinal focusing. The dashed curves show the angular distribution without bremsstrahlung. The laser intensity parameter $\xi=70$ is kept constant. In panel (b) the normalized differential number for the focused laser field without longitudinal focusing is illustrated for different beam waist sizes $w_0$, where the pulse energy $W_G\propto \xi^2 w_0^2$ is kept constant. The legend shows the corresponding diffraction angles and laser intensity parameters. \label{Fig6}}
	\end{figure}
	
	Fig.~\ref{Fig6}a shows the normalized angular distribution $dN/d\vartheta$ of the created electrons, comparing focused and plane-wave laser pulses. We analyze both, the case driven by a single $\gamma$ photon and the scenario in which the bremsstrahlung spectrum is included. The blue solid line shows the distribution for the focused laser field without longitudinal focusing. The maximum of this distribution forms a characteristic peak shape located at about $\vartheta_0=0.3$\,mrad. The distribution remains practically unchanged when the longitudinal focusing is included (blue dotted line). Both distributions differ only minimally in absolute height (approx. 1\%), but not in the shape of the distribution. The blue dashed curve shows the distribution with bremsstrahlung for the focused pulse with a single $\gamma$-photon. The maximum shifts to smaller angles and lies at $\vartheta_0\approx0.2$\,mrad. The width decreases significantly because for a single $\gamma$-photon the ratio between the energy of the $\gamma$-quantum and the absorbed laser photons is the largest. The red curves show the spectrum for the pulsed plane wave. Here, the solid curve indicates the distribution including bremsstrahlung and the dashed curve for a single $\gamma$-photon. Compared to the focussed case, the distributions here are much broader. This is due to the fact that in the case of the pulsed plane wave the intensity in the interaction volume is the same, whereas in the case of focusing, the intensity decreases towards the edges. However, a larger intensity leads to the fact that more photons are absorbed from the laser field whereby also the transversal momentum becomes larger and accordingly the spectrum becomes broader.  Also in case of the pulsed plane wave, omitting the bremsstrahlung leads to a more narrow distribution. Overall, the emission angles are close to the forward direction ($\vartheta = 0$) because the energy of the bremsstrahlung photons is much larger than the energy absorbed from the counterpropagating laser wave.
		
	In  Fig.~\ref{Fig6}b, we compare the normalized angular spectrum including bremsstrahlung for different focusing strengths at constant laser pulse energy $W_G\propto \xi^2 w_0^2$. For reference, we compare $w_0 =2\,\mu m$ with the values $w_0 =1$\,$\mu$m and $3$\,$\mu$m, respectively, with the diffraction angles $\epsilon_L = 0.1275, 0.2551$ and $0.085$. The differential number behaves in such a way that the weaker focusing leads to a distribution that shifts the maximum from $\vartheta \approx 0.3$\,mrad towards smaller angles ($\vartheta_0=0.2$\,mrad) compared to the initial choice of parameters. In contrast, a stronger focusing leads to a broader distribution and a slight shift of the maximum towards bigger angles, because as the laser intensity increases, the transverse particle momentum grows and the width of the angular distribution is enhanced.	
	
\begin{figure}[h]
		\centering
		\includegraphics[width=1\linewidth]{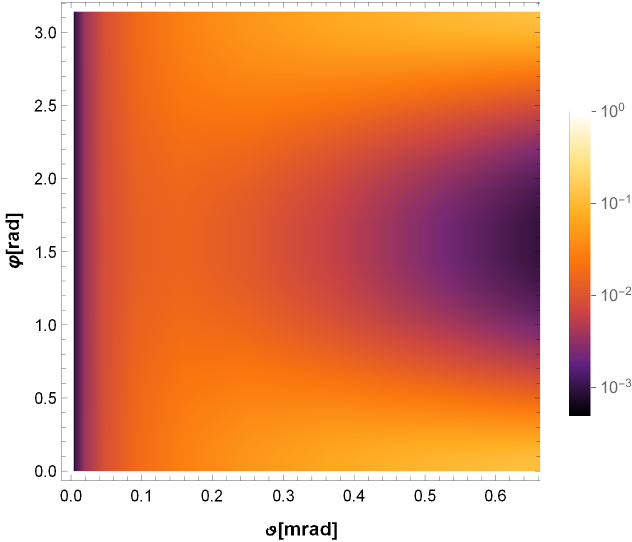}
		\caption{The normalized angular distribution $\frac{\text{d}^2N}{\text{d}\vartheta\text{d}\varphi}$ including bremsstrahlung and transversal focusing is shown as a density plot depending on $\varphi$ and $\vartheta$. The magnitude of the number is colour-coded according to the color bar on the right side.\label{Fig7}}
	\end{figure}

	In all angular distributions, the peak-shaped maximum at small angles is very peculiar. If we look at the twofold angular distribution function $\frac{\text{d}^2N}{\text{d}\vartheta\text{d}\varphi}$, keeping the azimuthal angle (measured with respect to the laser polarization direction) fixed at $\varphi=0$, the maximum would shift significantly and be around $\vartheta_0=1$\,mrad. For the azimuthal angle, the largest contributions come from $\varphi_0= 0, \pi, 2\pi$, but the intermediate angles for very small polar angles $\vartheta$ contribute just as much, while for increasing polar angles almost only very small areas around $\varphi_0$ contribute. This behaviour can be seen in Fig.~\ref{Fig7}. There, $\frac{\text{d}^2N}{\text{d}\vartheta\text{d}\varphi}$ including bremsstrahlung is shown as a color-coded density plot depending on $\varphi$ and $\vartheta$.
		
	We saw in Sec.~\ref{SWRRE} that radiation reaction effects exert a pronounced impact on the energy spectra of the created particles. In principle, also the angular spectra can be modified by the radiation reaction. However, it was shown by simulations in Ref.~\cite{X2} that the semiclassical model of Sec.~\ref{RRSEC} is not well suited to this end. By considering electrons of around 1\,GeV energy colliding head-on with high-intensity circularly polarized laser pulses of $\xi = 100$ it was found that stochasticity effects need to be included to predict the final electron angular distributions. While simulations based on the LL equation yielded pronounced deflections to the side, a QED treatment led to aggregation of the electrons in forward direction. Since in our scenario the created particles are emitted under very small angles (see Fig.~\ref{Fig6}), we may therefore expect that the radiation reaction will not modify their angular spectra qualitatively. However, a more rigorous QED investigation similarly to Ref.~\cite{X2} would be required to quantify the influence of radiation reaction on the particles' angular distributions. Such an analysis is nonetheless beyond the scope of the present paper and shall be addressed in a future study.
	
	\section{Conclusion\label{conclusion}}
	
	Energy spectra and angular distributions of the particles created by the nonlinear Breit-Wheeler process in the nonperturbative regime have been calculated, taking effects from laser focusing, bremsstrahlung $\gamma$-photons with broad frequency spectrum, and radiation reaction into account.
	
	Considering parameters of an upcoming experiment, we have found that the energy spectra of the particles shift to slightly higher energies when the laser focusing is included, as compared with the plane-wave case. This effect is caused by the transverse focusing, whereas the longitudinal focusing has turned out to be immaterial. However, when radiation reaction on the particles in the strong laser pulse is taken into account, their energies---measurable at a detector---are shifted to substantially smaller values. This effect largely exceeds the moderate upshift due to laser focusing.
	
	The angular distribution has a pronounced peak at low angles in all cases. This is due to the fact that for small polar angles $\vartheta$ the complete width of azimuthal angles contributes to the differential number, while for larger $\vartheta$ only a small range of azimuthal angles contributes. Compared to the plane wave, the consideration of focusing leads to a narrower angular spectrum. With constant pulse energy, a stronger focus, by contrast, leads to a broadening of the spectrum. The inclusion of bremsstrahlung has broadened the spectrum in all cases. 
	
	In further studies it would be interesting to improve the description of radiation reaction effects by a QED treatment, which would also be applicable to the angular distributions of the created particles. A fundamental open question is, moreover, to which extent radiation reaction effects can directly affect the pair production step itself.

	\begin{acknowledgements}
	This work has been funded by the Deutsche Forschungsgemeinschaft (DFG) under Grant No. 392856280 within the Research Unit FOR 2783/1. We thank A. Di Piazza for useful     discussions and O. Mathiak for his help at the onset of this study.
	\end{acknowledgements}


\end{document}